\documentclass[runningheads]{llncs}
\usepackage{graphicx}
\usepackage{amssymb}
\usepackage[applemac]{inputenc}
\usepackage{epstopdf}
\begin{document}
 \title{Perceptive, non-linear Speech Processing and Spiking 
 Neural Networks}
 \titlerunning{Perceptive, non-linear Speech Processing and Spiking 
 Neural Networks}
 \toctitle{Perceptive Speech Processing and Spiking Neurones}
 \author{Jean~ROUAT \thanks{This work has been funded by NSERC, MRST of 
 Qu\'ebec gvt., Universit\'e
 de Sherbrooke and by Universit\'e du Qu\'ebec \`a Chicoutimi. Many
 thanks to 
 Peter Murphy and
  Daniel Pressnitzer for proofreading this paper and to our COST277 
  collaborators~: Christian Feldbauer and Gernot 
 Kubin from Graz University of Technology
 for fruitful 
 discussions on analysis/synthesis filterbanks%
 .}, Ramin~PICHEVAR and Stéphane~LOISELLE 
 }
 \authorrunning{J.~Rouat, R.~Pichevar and S.~Loiselle}
 \tocauthor{Rouat, Pichevar and Loiselle}
 \institute{Universit\'e de Sherbrooke \\
 \email{http://www.gel.usherbrooke.ca/rouat/}
 }
 \maketitle
\textbf{Preprint of paper:}\\
Rouat, J., Pichevar, R., Loiselle, S. (2005). Perceptive, Non-linear Speech Processing and Spiking Neural Networks. In: \textit{Chollet, G., Esposito, A., Faundez-Zanuy, M., Marinaro, M. (eds) Nonlinear Speech Modeling and Applications.} NN 2004. Lecture Notes in Computer Science, vol 3445. Springer, Berlin, Heidelberg. https://doi.org/10.1007/11520153\_14
%

  \begin{abstract}
 Source separation and speech recognition are very difficult in 
 the context of noisy and corrupted speech. Most conventional 
 techniques need huge databases to estimate speech (or noise) density 
 probabilities to perform separation or recognition. We discuss the 
 potential of perceptive speech analysis and processing in combination 
 with biologically plausible neural network processors. We illustrate 
 the potential of such non-linear processing of speech
on a source separation system inspired by an Auditory 
 Scene Analysis paradigm.
  We also discuss a potential application in speech recognition.
  \end{abstract}
  \textbf{keywords:}
  Auditory modelling, Source separation, Amplitude Modulation, Auditory Scene
  Analysis, Spiking Neurones, Temporal Correlation, Multiplicative
  Synapses, Cochlear Nucleus, Corrupted Speech Processing, Rank Order
  Coding, Speech recognition.




  \section{Introduction}
  Processing of corrupted speech is an important research field with many
    applications in speech coding, transmission, recognition and audio 
    processing. For example, speech enhancement, auditory modelling and source separation
   can be used to
    assist robots in segregating multiple speakers, to ease the
    automatic transcription of videos via the audio tracks, to segregate
    musical instruments before automatic transcription, to clean up a
    signal before performing speech recognition, etc. The ideal
    instrumental set-up is based on the use of arrays of microphones during
    recording to obtain many audio channels.
  In that situation,
  very good separation can be
   obtained between noise and the signal of
   interest~\cite{widrow1975,kaneda1986,hyvarinen2001} and experiments with
   good
   enhancement have been reported in speech
   recognition
   ~\cite{vancompernolle1990,seltzer2002} 
   \cite{brandstein2001,haykin2002}~\cite{sanchez-bote2003}.
  Applications have been ported on mobile
  robots~\cite{nakadai2002}
  ~\cite{valin2003,valin2004a} and have also been developped to track  
  multiple speakers
  ~\cite{potamitis2003}.

 In many situations, only one channel is available to the audio
   engineer that still has to clean the signal and solve the separation problem.
    The cleaning of corrupted speech is, then, much more difficult.
    From the scientific literature, most of the proposed monophonic systems perform
    reasonably well on specific signals (generally voiced speech) but fail to efficiently
    segregate a broader range of signals. These relatively negative
    results may be overcome by combining and exchanging expertise and knowledge
    between engineering, psycho-acoustic, physiology and computer 
    science. Statistical approaches like Bayesian networks,
    Hidden Markov Models and one microphone ICA perform reasonably well
    once the 
    training dataset or the probability distributions have been suitably 
    estimated. But these approaches usually require supervised training on huge 
    databases and are designed for specific applications. 
    On the other hand, perceptive and bio-inspired approaches require less training,
    can be unsupervised and offer strong potential even if they are less mature.
   In the present work we are interested in monophonic bio-inspired 
   corrupted-speech processing
   approaches. 
  
    \section{Perceptive Approach}
    We propose to combine knowledge from psycho-acoustics, psychology 
	and neurophysiology to propose new non-linear processing systems for 
	corrupted speech. From physiology we learn that the auditory 
	system extracts simultaneous features from the underlying signal, 
	giving birth to simultaneous multi-representation of speech. We 
	also learn that fast and slow efferences can selectively enhance speech 
	representations in relation to the auditory environment.
	This is in opposition with most conventional speech processing systems 
	that use a
    systematic analysis~\footnote{Systematic analysis extracts the
    same features independently of signal context. Frame by frame 
    extraction of Mel Frequency
    Cepstrum Coefficients (MFCC) is an example  of systematic
    analysis.}
    that is effective only when speech segments under
    the analysis window are
     relatively stationary and stable.
       
	Psychology observes and attempts to explain the auditory 
	sensations by proposing models of  hearing. The interaction 
	between sounds and their perception can be interpreted in terms 
	of auditory 
	environment or auditory scene analysis. We also learn from 
	psycho-acoustics that the time structure organisation of speech and sounds 
	is crucial in perception. In combination with physiology, suitable 
	hearing models
	can also be 
	derived from research in psycho-acoustic.

    \subsection{Physiology: Multiple Features}

 Inner and outer hair cells establish synapses with efferent and 
    afferent fibres. The efferent projections to the inner hair cells 
    synapse on the afferent connection, suggesting a modulation of the 
    afferent information by the efferent system. On the contrary, other 
    efferent fibres project directly to the outer hair cells, 
    suggesting a direct control of the outer hair cells by the 
    efferences. It has also been observed that all afferent fibres (inner 
    and outer hair cells) project directly into the cochlear nucleus. 
    The cochlear nucleus has a layered structure that preserves 
    frequency tonotopic organisation. One finds very different 
    neurones that respond to various features~\footnote{onset, 
    chopper, primary-like, etc.}. 
    Schreiner and Langner~\cite{schreiner1986,schreiner1988} have shown that
    the inferior colliculus of the cat contains a highly systematic
    topographic representation of AM parameters. Maps showing best
    modulation frequency have been determined. The pioneering work by Robles, Ruggero and
    Evans~\cite{robles1991}
    \cite{evans1992}\cite{ruggero1992}
    reveals the importance of AM-FM~\footnote{Other features like transients,
    ON, OFF responses are observed, but are not implemented in this 
    paper.}
    coding in the peripheral auditory system along with the role of the efferent
    system in relation with adaptive tuning of the cochlea. Recently, 
    small  neural circuits in relation with \emph{wideband 
    inhibitory input} neurones have been observed by Arnott
    \textit{et al.}~\cite{arnott2004} in the cochlear nucleus. These 
    circuits, explain the response of specialised neurones to
    frequency position of sharp spectral notches. Pressnitzer
    \textit{et al.}~\cite{pressnitzer2001} have proposed another use 
    for such networks for auditory scene analysis.

    It is also known that the auditory efferent system
    plays a crucial role in
    enhancing signals in background 
    noise~\cite{Giguere94}~\cite{liberman1996}~\cite{kim2002}. 
    Kim \textit{et al.}~\cite{kim2002} measure the effect of aging on the medial 
    olivocochlear system and suggest that the
    functional decline of the medial 
    olivocochlear system with age precedes
    outer hair cell degeneration.
    
    It is clear from physiology that multiple and simultaneous representations
       of the same input
       signal are observed in the cochlear
       nucleus~\cite{henkel1997}~\cite{tang1996}. In the remaining parts of the
       paper, we call these representations, \emph{auditory images}. 
       It is interesting to note that Harding and Meyer \cite{harding2003} propose a
     multi-resolution Auditory Scene Analysis
	that uses both high- and low- resolution representations
	of the auditory signal in parallel.

    \subsection{Cocktail-party Effect and Auditory Scene Analysis}
    Humans are able to segregate a desired source in a mixture of
	sounds (\emph{cocktail-party effect}). Psycho-acoustical experiments have shown that
	although binaural audition may help to improve segregation
	performance, human beings are capable of doing the segregation
	even with one ear or when all sources come from the same
	spatial location (for instance, when someone  listens to a radio
	broadcast)~\cite{bregman1994}.
	Using the knowledge acquired in visual scene analysis
	and by making an analogy between vision and audition,
	Bregman developed the key notions of \textit{Auditory Scene 
	Analysis} (ASA)%
	~\cite{bregman1994}. Two of the most important aspects in  ASA are the
	\emph{segregation} and \emph{grouping (or integration)} of sound sources.
	The segregation
	step partitions the auditory scene into  fundamental auditory
	elements and the grouping is the binding of these elements in
	order to reproduce the initial sound sources. These two stages are
	influenced by top-down processing (schema-driven). The aim in
	Computational Auditory Scene Analysis (CASA) is to develop
	computerised methods for solving the sound segregation problem by
	using psycho-acoustical cues and physiological characteristics~\cite{beauvois1991,%
	wang1999}. For a review see~\cite{cooke2001a}.

\section{Spiking Neural Networks}%
\label{sec:spikeNN}
The previous section introduced the perceptual approach to non-linear 
speech analysis. In this section we emphasise the computational power
of spiking neurones in the context of speech and signal processing.
Autonomous bio--inspired and spiking neural networks are an
alternative to supervised systems. A good review on bio-inspired
spiking neurones can be found in \cite{handbookNNComputation1997}
and in books such as~\cite{Maass98}, \cite{wermter2001} and
\cite{Gerstner2002}.

\subsection{Analysis and Perception}
In perception, the
recognition of stimuli is quasi-instantaneous,
even if the
information propagation speed in living neurones is slow. This
phenomenon is well documented for hearing and visual
systems~\cite{deweese2002}~\cite{thorpe1996,thorpe2001}. It
implies that neural responses are conditioned by previous events
and states of the neural sub-network~\cite{natschlager2003}. The
understanding of the underlying mechanisms of perception in
combination with that of the peripheral auditory
system~\cite{delgutte1980} \cite{frisina1985} \cite{popper1992}
\cite{hewitt1994} 
\cite{zotkin2003} helps the researcher in designing speech analysis
modules.

\subsection{Intuitive Notion of Spiking Neurones}
\begin{figure}[]
\centering
\includegraphics[width=5cm]{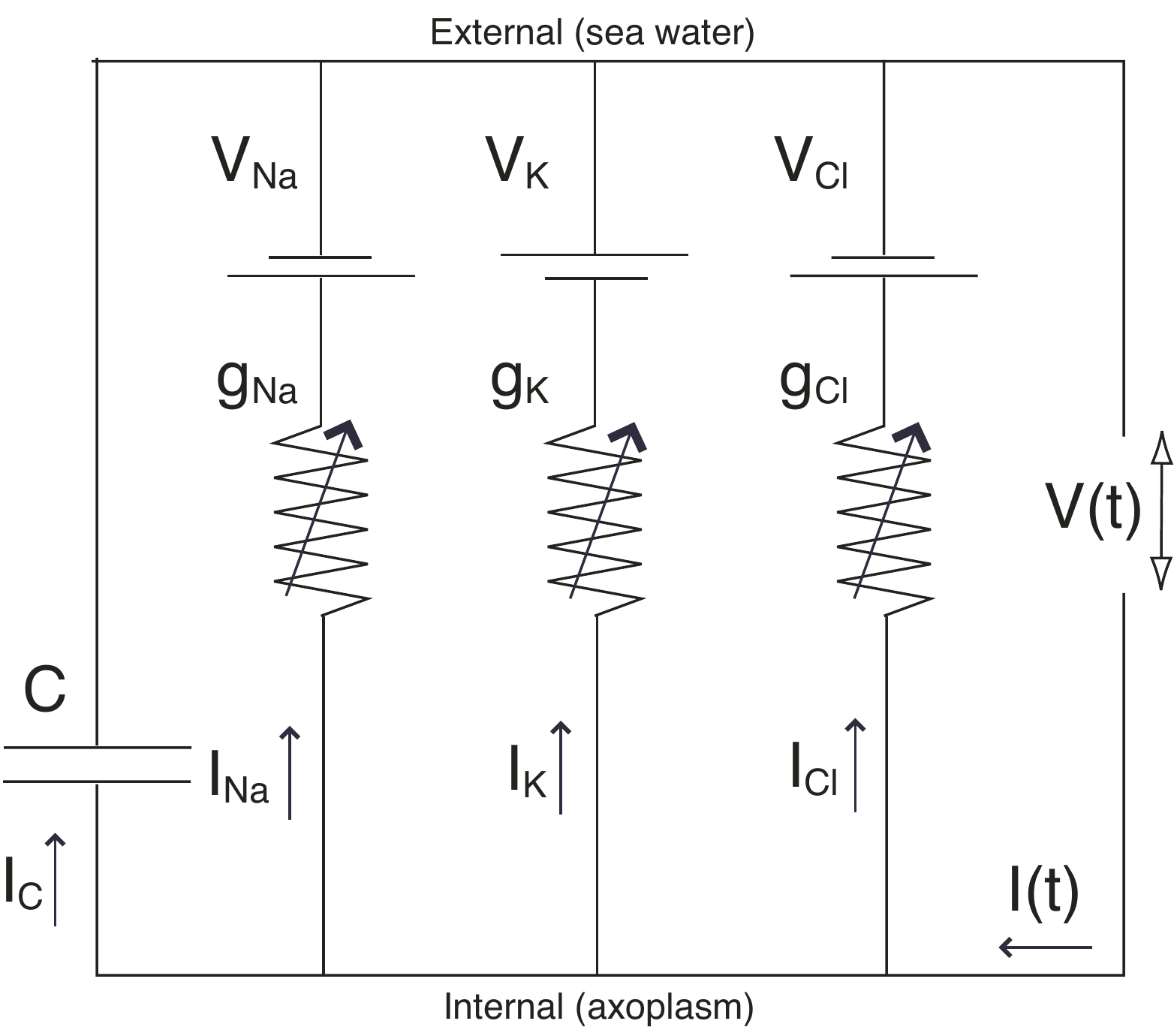}
\caption{Equivalent circuit of a membrane section of the giant squid axon 
(from Hodgkin-Huxley, 1952). $g_{Cl}$, $g_{Na}$ and $g_{K}$ are the 
conductance of the membrane for respective ionic gates. $V(t)$ is the membrane 
potential when $I(t)=0$ (no external input).}
\label{fig:hh}
\end{figure}
\begin{figure}
   \centering
 \includegraphics[width=3cm]{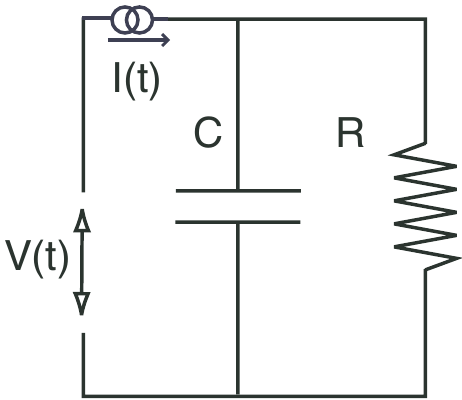}
 \caption{Equivalent circuit of a leaky integrate and fire neurone. 
 C: membrane capacitance, R: membrane resistance, V: membrane 
 potential.}
 \label{fig:LIF}
 \end{figure}

In the case of bio-inspired neural networks, temporal sequence processing 
is done naturally
because of the intrinsic dynamic behaviour of neurones. The pioneering work
in the field
of neural networks has been done by Hodgkin and Huxley 
(H\& H) at the University
of Plymouth, who proposed a mathematical description of the behaviour of 
the giant
squid axon. Although this model is complete so far (it can predict most of the
behaviours seen in simple biological neurones), it is very complex and 
difficult to simulate
in an artificial neural network paradigm. A very simplified model of the H\& 
H is the Leaky Integrate and Fire model (LIF) as presented in 
figure~\ref{fig:LIF}.

%
 
 $I(t)$ is the sum of the current going 
 trough the capacitance plus 
 the resistance current. The sub-threshold~\footnote{subthreshold 
 potential assumes that the internal potential V(t) is sufficiently
 small in comparison to the
 neurone's internal
 threshold. Supra-threshold potential is the potential of the 
 neurone when it is greater than the internal threshold. The 
 Hodgkin-Huxley model integrates 
 sub-threshold, threshold and supra-threshold activity in the same set 
 of equations. It is not the case with the LIF model.
 } potential $V(t)$ is given 
 by:
 
 \begin{equation}
 I(t) = C(t) \frac{dV(t)}{dt} + \frac{V(t)}{R(t)}
 \label{eq:I}
 \end{equation}
 
 $V(t)$ is the output, $I(t)$ is the input. When $V(t)$ crosses a 
 predetermined
 threshold $\delta(t)$, the neuron fires and emits a spike. Then 
 $V(t)$ is reset to $V_{r}$, where $V_{r}$ is the resting potential. 
 
In this paper we use 
a 
non-linear oscillator to reproduce the behaviour of
the Integrate and Fire neurone.

In the following subsections, we review some of the studies that are pertinent to speech
processing and source separation in spiking neurones.

\subsection{Formalisation of Bio-inspired Neural Networks}
There are many publications which describe mathematical formalism for
spiking neurones. One can cite for example the work by
Maass~\cite{maass1997b} and his team, in which they have shown that
networks of spiking neurones are computationally more powerful
than the models based on McCulloch Pitts neurones. In
\cite{natschlager2003} the authors have shown that information about the
result of the computation is already present in the current neural
network state long before the complete spatio-temporal input
patterns have been received by the neural network. This suggests
that neural networks use the temporal order of the first spikes
yielding ultra-rapid computation, according to the observations
by~\cite{thorpe1996}. In \cite{maass2000} and
\cite{natschlager2001}, the authors explain how neural networks
and dynamic synapses (including facilitation and depression) are
equivalent to a given quadratic filter that can, thus, be approximated by
a small neural system. The authors also show that any filter that can be
characterised by a Volterra series can be approximated with a single
layer of spiking neurones.


\subsubsection{Mutual information and Pattern recognition.}
Among the many publications in information theory one can cite the
works by Fred Rieke \textit{et al.}~\cite{rieke1997}, 
Sejnowski~\cite{sejnowski1995}, DeWeese~\cite{deWeese1996} and Chechik 
and Tishby~\cite{chechik2000}
where it is shown that spike coding in neurones is close to
optimal and that plasticity in Hebbian learning rule increases
mutual information close to optimal transmission of information.

\subsubsection{Novelty detection.}
For unsupervised systems, novelty detection is an important property
that facilitates autonomy (robots can detect if stimuli is new or
already seen). When associated with conditioning, novelty detection
can create autonomy of the
system~\cite{ho1998a}~\cite{Borisyuk2001}.

\subsubsection{Sequence classification.}
Sequence classification is particularly interesting for speech.
Recently Panchev and Wermter~\cite{panchev2003} have shown that synaptic plasticity can be
used to perform recognition of sequences. Perrinet~\cite{perrinet2003}
and Thorpe~\cite{thorpe2001} discuss
the importance of sparse coding and rank order coding for
classification of sequences.

\subsection{Segregation and Integration with Binding}
Neurone assemblies (groups) of spiking neurones can be used to 
implement segregation and fusion (integration) of objects in an
auditory image representation. Usually, in signal processing,
correlations (or distances) between signals are implemented with
delay lines, products and summation. 
With spiking neurones, comparison (temporal correlation) between
signals can be made without implementation
of delay lines.
In section~\ref{sec:SourceSep}, page~\pageref{sec:SourceSep}, we use
that approach
 by presenting auditory images to spiking
neurones with dynamic synapses. Then, a spontaneous organisation
appears in the network with sets of neurones firing in synchrony.
Neurones with the same firing phase belong to the same auditory
objects. In 1976 and 1981, the temporal correlation that performs
binding was proposed by Milner~\cite{milner1974}
and independently by Malsburg~\cite{malsburg1981,malsburg1986,malsburg1999}.
 Milner and Malsburg have observed that synchrony is a crucial
 feature to bind neurones that are associated with similar characteristics.
 Objects belonging to the same entity
 are bound together in time.  In other words, synchronisation
 between different neurones and de-synchronisation among different
 regions perform the binding.
 To a certain extent, this property has been exploited by
Bohte~\cite{bohte2002} to perform unsupervised clustering for
recognition on images, by Schwartz~\cite{schwartz1989} for vowel processing
with spike synchrony between cochlear channels, by Hopfield~\cite{Hopfield1995} 
to propose pattern recognition with spiking
neurones, by Levy \textit{et al.}~\cite{levy2001} to perform cell assembly of spiking
neurones using Hebbian learning with depression. Wang and Terman~\cite{wang1997}
have proposed an efficient and robust technique for image
segmentation and study its potential in CASA (Computational Auditory 
Scene Analysis)~\cite{wang1999}.

\subsection{Example of Associative Neural Network}
 Alkon \textit{et al}.~\cite{alkon1990} 
   have shown that dendrites can learn associations between input sequences
   without knowledge about neurone outputs. They derive an image
   recognition application~\cite{blackwell1992} from this work. The  network
   associatively learns correlation and anti-correlation between time
   events occurring in pre-synaptic neurones.  Those neurones synapse on the
   same element (same area) of a common post synaptic neurone.  A learning
   rule modifies the cellular excitability at dendritic patches.  These
   synaptic patches are postulated to be formed on branches of the
   dendritic tree of vertebrate neurones.  Weights are
   associated to patches rather than to incoming connection.  After
   learning, each patch characterises a pattern of activity on the input
   neurones.
    In comparison with
   most commonly used networks, the weights are not used to store the
   patterns and the comparison between patterns is based on 
   normalised correlation
   instead of projections between the network input vectors and the neurone
   weights.  Based on this type of network, a prototype vowel recognition system
   has been designed and preliminary results have shown that the short-time AM
   structure carries information that can be used for recognition of
   voiced speech~\cite{rouat1998}. One of the main drawbacks of that
   approach is that explicit encoding of
    reference patterns in dendritic patches of neurones is required.
   
\subsection{Rank Order Coding}
Rank Order Coding  has been 
proposed by Simon Thorpe and his team from CERCO, Toulouse to explain 
the impressive performance of our visual 
system~\cite{rvanrullenvision2002,perrinet2003}.
The information is distributed through 
a large population of neurones and 
is represented by spikes relative timing
in a single wave of action potentials.
The quantity of information that can be transmitted by this type of code increases
with the number of neurones in the population.
For a relatively large number of neurones, the code transmission 
power can satisfy the needs of any visual task~\cite{rvanrullenvision2002}.
There are advantages in using the relative order and not the exact spike latency:
the strategy is easier to implement, the system is less subject to changes
in intensity of the stimulus and the information is available 
as soon as the first spike is generated.

%


\subsection{Summary}

Bio-inspired neuronal networks are  well adapted to signal
processing where time is important. They 
can be fully unsupervised. Adaptive and unsupervised
recognition of sequences is a crucial property of living neurones.
Among the many properties we listed in this section, the paper
implements the segregation and integration with sets of
synchronous neurones. 
At the moment, this work does not reflect the
full potential of spiking neurones and is more or less
exploratory.

  \section{Source Separation}\label{sec:SourceSep}

%
  Most monophonic source separation systems require \textit{a priori} 
  knowledge, i.e. expert systems (explicit knowledge) or
  statistical approaches
  (implicit
  knowledge)~\cite{cooke2001a}.
  Most  of these systems perform
   reasonably well only on specific signals (generally voiced speech or 
   harmonic music)
   and fail to efficiently
   segregate a broad range of signals.
    Sameti~\cite{sameti1998} uses Hidden Markov Models, while
    Roweis~\cite{roweis2000,roweis2003},
    and Gomez~\cite{gomez2003}
    use Factorial Hidden Markov Models. Jang and Lee~\cite {jang2003}
    use  Maximum A Posteriori (MAP) estimation. They all require
    training on huge signal databases to
    estimate probability models.
  Wang and Brown \cite{wang1999} 
  proposed an original bio--inspired approach that uses features 
  obtained from
  correlograms and F0 (pitch frequency) in combination with an oscillatory
  neural network.
  Hu and Wang use a pitch tracking technique
   \cite{hu2004} to segregate harmonic sources.
 Both systems are limited  to harmonic signals.

 We propose here to extend the bio-inspired approach to more general 
 situations without training or prior
    knowledge of underlying signal properties~\footnote{Prior 
    knowledge is embodied in the representations of the acoustic 
    signals.}.

    \subsection{Binding of Auditory Sources }\label{examples}
       Various features of speech are extracted in different areas of 
       the brain~\footnote{AM and FM maps are observed in the 
       colliculus of 
       the cat, onset neurones are present in the cochlear nucleus, etc.}.
       We assume here that sound segregation is a generalised
       classification problem, in which we want to bind features
       extracted from the auditory image representations
       in different regions of our neural network map.
    \subsection{System Overview}
    \begin{figure}[h]
    \centering
     \includegraphics[width=10cm] {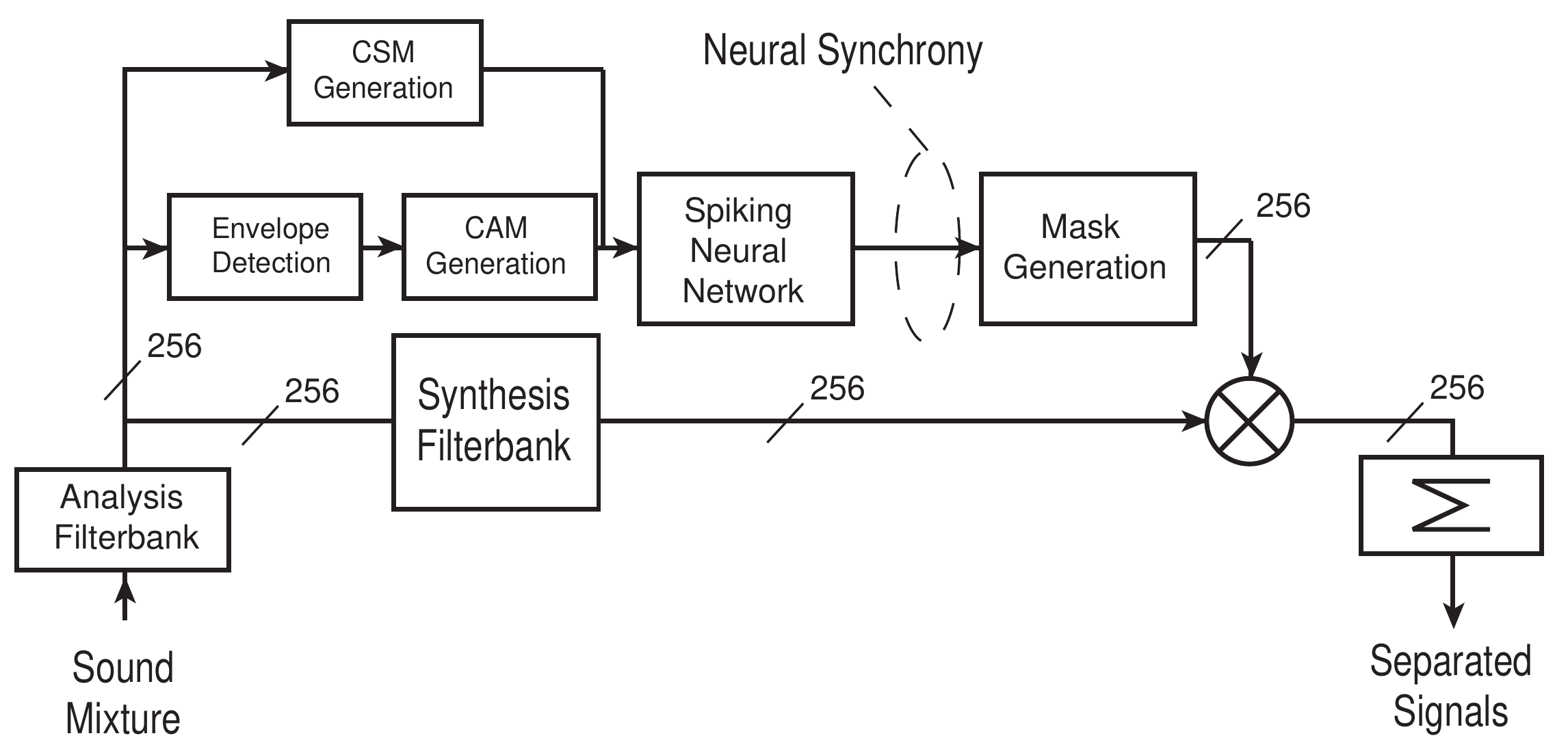}
    \caption{Source Separation System. Depending on the sources auditory 
    images (CAM or CSM), the spiking neural network 
    generates the mask (binary gain)  to switch ON/OFF -- in time and
    across channels -- the synthesis 
    filterbank channels before final summation.}
    \label{fig:systemArchitecture}
    \end{figure}
  In this work
analysis and recognition are
integrated. 
  Physiology, psychoacoustic and
  signal processing are
  integrated
 to design a multiple sources separation system when only one audio
  channel is available  
 (Fig.~\ref{fig:systemArchitecture},
 page \pageref{fig:systemArchitecture}).
 It combines a
 reconstruction analysis/synthesis cochlear filterbank
 along with auditory image representations of audible signals with 
 a spiking neural network.
 The segregation and binding of
 the auditory objects (coming from different sound sources) is
 performed by the spiking neural network (implementing the \emph{temporal
 correlation}~\cite{milner1974,malsburg1981}) that also generates a
 mask~\footnote{Mask and masking refer here to a binary gain and 
 should not be confused with the conventional definition of masking in 
 psychoacoustics.} to be used in conjunction
 with the synthesis filterbank to generate the separated sound sources.

 The neural network 
 uses third generation
 neural networks, where  neurones are usually called 
 \textit{spiking} neurones~\cite{maass1997b}. 
 In our implementation,  neurones firing at the same instants (same firing phase)
 are characteristic of similar stimuli or comparable input signals%
 ~\footnote{The information
 is coded in the firing instants.}. \textit{Spiking}  neurones, in opposition 
 to \textit{formal}
  neurones, have a constant firing amplitude.
 This coding yields noise and interference robustness while facilitating 
 adaptive
 and dynamic
 synapses (links between  neurones) for unsupervised and autonomous
 system design. Numerous spike timing coding schemes are possible (and
 observable in physiology)~\cite{haines1997}. Among them, we decided to 
  use synchronisation and oscillatory
 coding schemes in combination with a competitive unsupervised framework
 (obtained with dynamic synapses),
 where groups of synchronous  neurones are observed.
 This choice has the advantage to allow design of unsupervised 
 systems with no training (or learning) phase. To some extent, 
 the neural network 
 can be viewed as a map where links 
 between  neurones are dynamic.
 In our implementation of the
 \emph{temporal correlation}, two  neurones with similar inputs 
 on their dendrites will
 increase their soma to soma synaptic weights (dynamic synapses), forcing synchronous
 response.
 On the opposite,  neurones with dissimilar dendritic inputs will have reduced
 soma to soma
 synaptic weights, yielding reduced coupling and asynchronous neural
 responses.

 \begin{figure}[h]
 \centering
  \includegraphics[width=6cm]{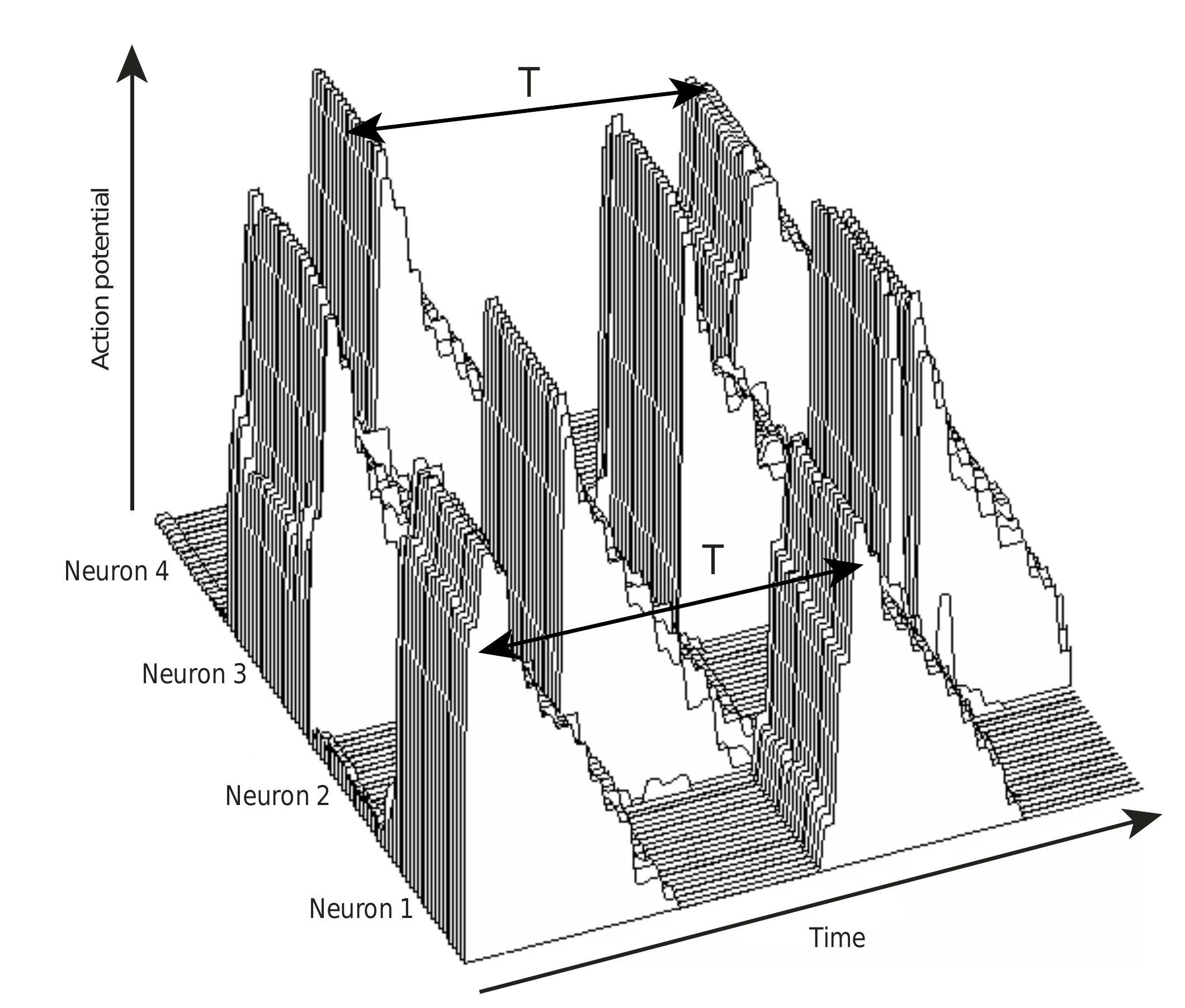}
 \caption{Dynamic temporal correlation for two
 simultaneous sources: time
 evolution of the supra--threshold electrical
 output potential for four neurones from the second layer (output 
 layer). T is the oscillatory period. Two sets of
 synchronous
 neurones appear (neurones 1 \& 3 for source 1; neurones 2 \& 4 for source 2). Plot
 degradations are due to JPEG 
 coding.}
 \label{fig:principeDLM}
 \end{figure}

 Figure~\ref{fig:principeDLM}, page~\pageref{fig:principeDLM}
 illustrates the oscillatory response behavior of the output layer of 
 the proposed neural network for two sources.

 
 While conventional signal processing
 computation of correlations encounter difficulties in taking
 simultaneously into
 account the spatial aspect (multi-step correlation has to be evaluated),
 the spiking neural network is able to compute a \textbf{spatio}-temporal 
 correlation
 of the input signals in one step.
 
 
  Compared to conventional approaches, our system does not
  require a priori knowledge, is not limited to harmonic signals,
  does not require training and does not need pitch extraction.
  %
 
  The architecture is also
  designed to handle continuous input signals (no need to segment
  the signal into time frames)  and is based  on the availability of simultaneous
  auditory representations of signals. 
 Our approach is inspired by knowledge in anthropomorphic systems but 
 is not an attempt to reproduce faithfully physiology or psychoacoustics.


   \subsection{Proposed System Strategy}\label{strategy}

  \begin{figure}[h]\centering
   \includegraphics [height=4cm, width=6cm]{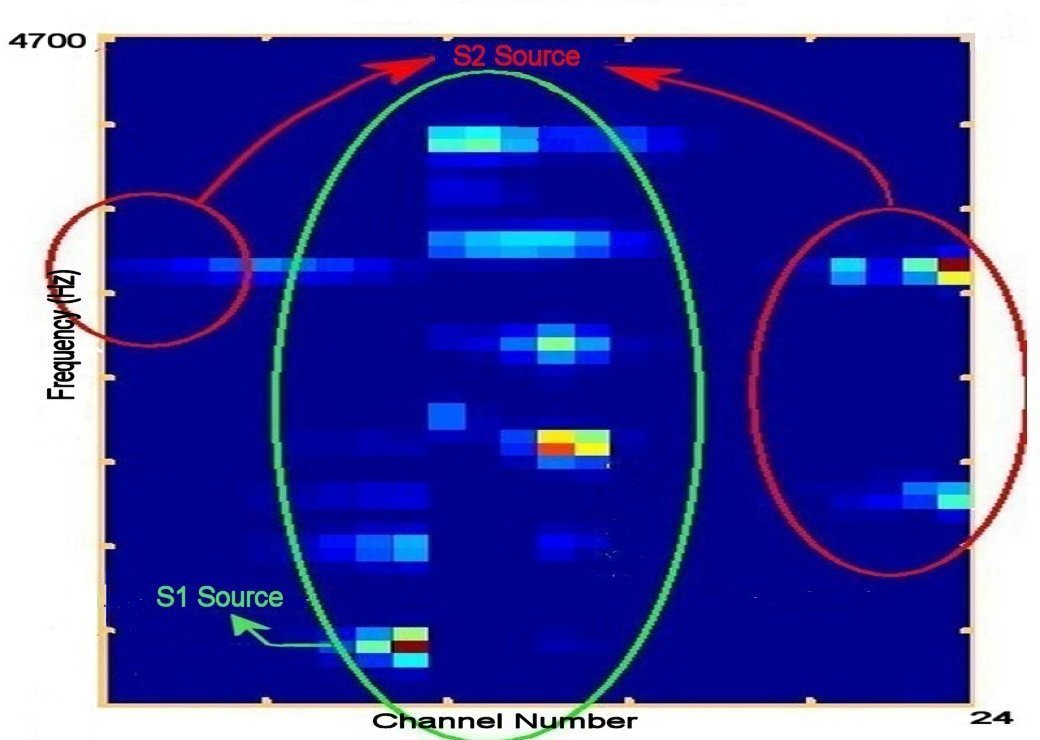} 
  \caption{Example of a twenty--four channels CAM for a mixture of /di/ and /da/
  pronounced by two speakers; mixture at $SNR= 0$ dB and frame center at
  $t= 166$~ms.
  }\label{fig:CAM}
  \end{figure}

   Two representations are simultaneously generated: 
   an amplitude modulation map, which
   we call Cochleotopic/AMtopic (CAM) Map~\footnote{To some extent, 
   it is  related to
   modulation spectrograms. See for example work by~\cite{atlas2003,Meyer2001}.}
  and
 %
   a Cochleotopic/Spec\-trotopic Map (CSM)  that encodes the averaged
   spectral energies of the cochlear filterbank output.
   The first representation somewhat reproduces the AM processing
   performed by multipolar cells (Chopper-S) from the anteroventral cochlear
   nucleus~\cite{tang1996}, while the second representation could be closer to
   the spherical bushy cell processing from ventral cochlear
   nucleus~\cite{henkel1997} areas.

  We assume that different sources  are disjoint in the
   auditory image representation space and that 
   masking (binary gain) of the
   undesired sources is feasible.
 Speech has a specific structure that
   is different from that of most noises and
   perturbations~\cite{rouat1997b}. Also, when dealing with simultaneous
   speakers, separation is possible when preserving the time structure
   (the probability at a given instant $t$ to observe overlap in pitch and
   timbre is relatively low). Therefore, a binary gain can be used to suppress the
   interference (or separate all sources with adaptive masks).

  \subsection{Detailed Description}\label{description}
  \subsubsection{Signal Analysis.} \label{preprocessing}
  Our CAM/CSM
    generation algorithm is as follows:
    \begin{enumerate}
    \item Down--sampling to 8000 samples/s.

    \item Filter the sound source using a 256-filter Bark-scaled
    cochlear filterbank ranging from $100$ Hz to $3.6$ kHz. 

    \item
    \begin{itemize}
    \item  For CAM: Extract the envelope (AM demodulation)
    for channels 30-256 (400--3600~Hz); for other low frequency channels (1--29: 
    100--400~Hz) use  raw
    outputs~\footnote{Low-frequency channels are said to resolve the 
    harmonics while others do not, suggesting a different strategy for 
    low frequency channels~\cite{rouat97}.}.

    \item For CSM: Nothing is done in this step.

    \end{itemize}

    \item Compute
    the STFT of the envelopes (CAM) or of the filterbank outputs (CSM)
    using a Hamming window~\footnote{Non-overlapping adjacent
    windows with 4ms or 32ms lengths have been tested.}.
    \item To increase the
    spectro-temporal resolution of the STFT, find the reassigned
    spectrum of the STFT~\cite{Plante98} (this consists of applying an
    affine transform to the points to re-allocate the spectrum).
    \item Compute the logarithm of
    the magnitude of the STFT. 
    The logarithm enhances the presence of the stronger source in a
    given 2D frequency bin of the CAM/CSM~\footnote{$\log(e_{1}+e_{2})
    \simeq max(\log \;e_{1},\log \;e_{2})$ (unless $e_{1}$ and $e_{2}$
    are both large and almost equal)~\cite{roweis2003}.}.

    \end{enumerate}

%
    For a given
    instant, depending on the signal, the peripheral auditory system
    can enhance the AM, the FM, the envelope, or the transient
    components of the signal.

  %
  %
%

  \subsubsection{The Neural Network.}\label{network}
  \paragraph{First layer: Image segmentation.}
  The dynamics of the neurones we use is governed by a modified
  version of the Van der Pol relaxation oscillator (Wang-Terman
  oscillators \cite{wang1999}). The state-space equations for these
  dynamics are as follows:
  \begin{equation}\label{neuronx}
   \frac{dx}{dt}= 3x-x^{3}+2-y+\rho+p+S
  \end{equation}
  \begin{equation}\label{neurony}
   \frac{dy}{dt}=\epsilon[\gamma(1+\tanh(x/\beta))-y]
  \end{equation}
  Where $x$ is the membrane potential (output) of the neurone and $y$
  is the state for channel activation or inactivation. $\rho$
  denotes the amplitude of  Gaussian noise, $p$ is the external
  input to the neurone, and $S$ is the coupling from other neurones
  (connections through synaptic weights). $\epsilon$, $\gamma$, and
  $\beta$ are constants~\footnote{In our simulation, $\epsilon = 0.02$, $\gamma=4$,
  $\beta=0.1$ and $\rho=0.02$.}. The Euler integration method is used to
  solve the equations.
  \begin{figure}[ht]\centering
   \includegraphics [height=6cm] {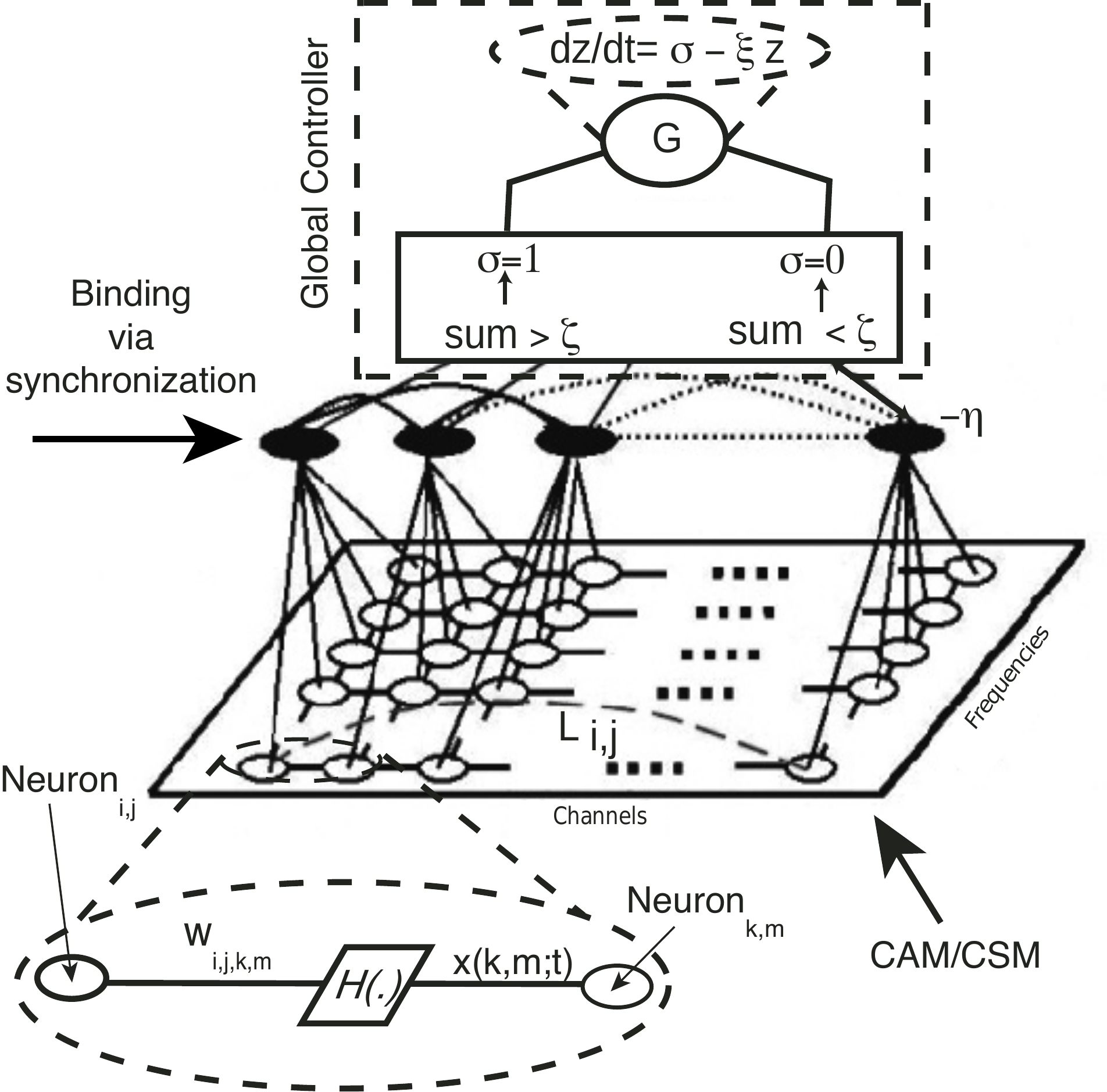}
  \caption{Architecture of the Two-Layer Bio-inspired Neural Network.
  G: Stands for global controller (the global controller for the
  first layer is not shown on the figure). One long range connection
  is shown. Parameters of the controller and of the input layer are 
  also illustrated in the zoomed areas.}\label{fig:networkenhancedWB}
  \end{figure}
  The first layer is a partially
  connected network of relaxation oscillators~\cite{wang1999}. Each
  neurone is connected to its four neighbors. The CAM (or the CSM) is
  applied to the input of the neurones. Since the map is sparse, the
  original 256 points computed for the FFT are down-sampled to 50
  points. Therefore, the first layer consists of 256 x 50 neurones.
 The geometric interpretation of pitch
  (ray distance criterion) is less clear for the first 29
  channels, where harmonics are usually resolved~\footnote{Envelopes 
  of resolved harmonics are nearly constants.}.
  For this reason, we have also established long-range
  connections from \emph{clear} (high frequency) zones to \emph{confusion}
  (low frequency) zones. These connections exist only across the
  \emph{cochlear channel number} axis in the CAM. 

  The  weight, $\ w_{i,j,k,m}(t)$ (figure~\ref{fig:networkenhancedWB}),
  between $neurone(i,j)$ and $neurone(k,m)$ of the first
  layer is:
  \begin{equation}\label{3}
   \ w_{i,j,k,m}(t)= \frac{1}{Card\{N(i,j)\}}\frac{0.25}{e^{\lambda|p(i,j;t)-p(k,m;t)|}}
  \end{equation}
  here $p(i,j)$ and $p(k,m)$ are respectively external inputs to
  $neurone(i,j)$ and $neurone(k,m)\in N(i,j)$. $Card\{N(i,j)\}$ is a
  normalization factor and is equal to the cardinal number (number
  of elements) of the set $N(i,j)$ containing neighbors connected to
  the $neurone(i,j)$ (can be equal to 4, 3 or 2 depending on the
  location of the neurone on the map, i.e. center, corner, etc.). The
  external input values are normalized. The value of $\lambda$
  depends on the dynamic range of the inputs and is set to
  $\lambda=1$ in our case. This same weight adaptation is used for
  \emph{long range clear to confusion zone} connections (Eq. \ref{long1})
  in the CAM processing case. The coupling $S_{i,j}$ defined in Eq.
  \ref{neuronx} is :
   \begin{equation}\label{influence}
   \ S_{i,j}(t)=\!\!\!\!\!\!\!\!\!\!\sum_{k,m\in
    N(i,j)}w_{i,j,k,m}(t)H(x(k,m;t))-\eta G(t)+\kappa L_{i,j}(t)
   \end{equation}
  $H(.)$ is the Heaviside function. The dynamics of $G(t)$ (the
  global controller) is as follows:
  \begin{equation}\label{intra1}
   G(t)=\alpha H(z-\theta)
  \end{equation}
  \begin{equation}\label{intra2}
   \frac{dz}{dt}=\sigma-\xi z
  \end{equation}
  $\sigma$ is equal to 1 if the global activity of the network is
  greater than a predefined $\zeta$ and is zero otherwise (Figure~\ref{fig:networkenhancedWB}).
  $\alpha$
  and $\xi$ are constants~\footnote{ $\zeta=0.2$, 
  $\alpha=-0.1$, $\xi=0.4$, $\eta = 0.05$ and $\theta=0.9$.}.

  $L_{i,j}(t)$ is the long range coupling as follows:
  \begin{equation}\label{long1}
  L_{i,j}(t)= \left\{ \begin{array}{lll}
	0 & j \geq 30 \\
	\sum_{k=225...256}
	w_{i,j,i,k}(t)H(x(i,k;t))& j<30 \end{array} \right.
  \end{equation}
  $\kappa$ is a binary variable defined as follows:
  \begin{equation}\label{long2}
  \kappa= \left\{ \begin{array}{lll}
      1 & for \qquad CAM\\0 & for \qquad CSM \end{array} \right.
  \end{equation}

  \paragraph{Second Layer: Temporal Correlation and Multiplicative
  Synapses.}
  The second layer is an array of 256 neurones
  (one for each channel). Each neurone receives the weighted
  product of
  the outputs of the first layer neurones along the frequency axis of
  the CAM/CSM.
  %
For the CAM: Since the geometric (Euclidian)
  distance between rays (spectral maxima) is a function of the pitch
  of the dominant source in a given channel, the weighted sum of the
  outputs of the first layers along the frequency axis tells us
  about the origin of the signal present in that channel. For
  the CSM: Highly localized energy bursts will be enhanced by that
  representation.
  Weights between layer one and layer two are defined as
  $w_{ll}(i)=\frac{\alpha}{i}$, where $i$ can be related to the
  frequency bins of the STFT and $\alpha$ is a constant for the CAM
  case, since we are looking for structured patterns. For the CSM,
  $w_{ll}(i)=\alpha$ is constant along the frequency bins as we are
  looking for energy bursts~\footnote{In our simulation, $\alpha=1$}.

  Therefore, the input stimulus to
  $neurone(j)$ in the second layer is defined as follows:
  \begin{equation}\label{secondweight2}
   \theta(j;t)=\overline{\prod_{i}w_{ll}(i)\Xi\{x(i,j;t)\}}
  \end{equation}
  The operator $\Xi$ is defined as:
  \begin{equation}\label{operator}
  \Xi\{x(i,j;t)\}= \left\{ \begin{array}{lll}
     1 & for \qquad x(i,j;t) =0\\ x(i,j;t) &  elsewhere \end{array}
  \right.
  \end{equation}
  where $\overline{()}$ is the  \textit{averaging over a time window} operator (the
  duration of the window is on the order of the discharge period).
  Multiplication is carried out only for non-zero outputs
  (in which spike is present) \cite{Gabiani2002,Pena2001}. A 
  functional analogue of this
  behavior has been observed in the integration of ITD (Interaural
  Time Difference) and ILD (Inter Level Difference) information in
  the barn owl's auditory system \cite{Gabiani2002} or in the
  monkey's posterior parietal lobe neurones that show
  \textit{receptive
   fields}
  that can be explained by a multiplication of retinal and eye or
  head position signals \cite{Andersen97}.

  The synaptic weights inside the
  second layer are adjusted through the following rule:
  \begin{equation}\label{1}
    \ w'_{ij}(t)= \frac{0.2}{e^{\mu|p(j;t)-p(k;t)|}}
  \end{equation}
   $\mu$ is chosen to be equal to $2$. The \emph{binding} of these features is 
   achieved via this
  second layer. In fact, the second layer is an array of fully
  connected neurones along with a global controller.  The dynamics of the second layer
  is given by an equation similar to equation~\ref{influence} 
  (without long range coupling). The global
  controller desynchronizes the synchronized neurones for the first
  and second sources by emitting inhibitory activities whenever
  there is an activity (spikings) in the network \cite{wang1999}.

  The selection
  strategy at the output of the second layer is based on temporal
  correlation: 
	   neurones belonging to the same
  source synchronize (same spiking phase) and neurones belonging to
  other sources desynchronize (different spiking phase).

  \subsubsection{Masking and Synthesis.}
  
  Time-reversed outputs of the \emph{analysis} filterbank are passed through
     the \emph{synthesis} filterbank giving birth to $z_{i}(t)$.
  Based on an output signal continuity criterion and on the phase synchronisation described in the previous
  section, a mask is generated by associating zeros and ones to
  different channels.
  \begin{equation}
       s(t)=\sum_{i=1}^{256}m_{i}(t) z_{i}(t)
       \label{eq:synthese}
   \end{equation}
   where $s(N-t)$ is the recovered signal (N is the length of the signal in discrete mode),
   $z_{i}(t)$ is the synthesis filterbank  output  for channel $i$ and
    $m_{i}(t)$ is the mask value.
  Energy is normalised in order to have the same SPL for all
  frames.
  Note that two-source mixtures are 
  considered throughout this article but the technique can be
  potentially used for more sources. In that case, for each
  time frame $n$, labeling of individual channels is  equivalent to
  the use of multiple masks (one for each source).

  \subsubsection{Experiments and Results.}\label{experiments}  
  Results can be heard and evaluated on one of the authors' web page:
  \cite{raminwww}~\cite{rouatwww}. Detailed results and experiments 
  are described in a companion paper by Pichevar and Rouat in the same 
  book where a comparison is made to the work by Hu~\cite{hu2004}, 
  Wang~\cite{wang1999} and Jang~\cite{jang2003}.

   Based on evidences regarding the dynamics of the efferent
   loops
   and on the richness of the representations
   observed in the Cochlear Nucleus, we propose a technique to
   explore the monophonic source separation problem using a
   multirepresentation bio-inspired pre-processing stage
   and a bio-inspired neural network that does not require any a priori
   knowledge of the signal.

 Results obtained from signal synthesis are
 encouraging and we believe that spiking neural networks in
 combination with suitable signal representations have a strong
 potential for use in speech and audio processing. The evaluation scores show 
 that the system yields 
 performance levels roughly comparable with other methods, to which it has been compared. 
 Furthermore, our method does not need any prior knowledge 
 and is not limited to harmonic signals.

 \section{Exploration in Speech Recognition}
 We illustrate here another  application in speech 
 recognition where perceptive signal analysis combined with 
 non-linear signal processing and spiking neural networks offers a 
 strong potential.
 
 \subsection{Auditory Based Features and Pattern Recognisers }
    Starting in the middle of the '80s, many auditory models have been
    proposed
    ~\cite{hunt1987,hunt1988},\cite{seneff1988},
    \cite{mceachern1992,mceachern1994},
    \cite{immerseel1993},\cite{ghitza1994,sandhu1995},
     \cite{slaney1990,slaney1993}, \cite{ainsworth1993}
    \cite{patterson1994,patterson1994a} and have been tested on
    speech processing systems. At that time, it was objected that
    auditory models were CPU-intensive and were only useful  when
    speech was noisy.

    Furthermore, that first generation of auditory-based models failed to be used
    by speech recognisers, since recognisers were not able to exploit the
    great granularity of auditory-inspired model outputs that preserve time structure and generate many simultaneous
    representations and features suitable to source separation and
    speech recognition. Furthermore, at  that  time, pattern recognisers were
    not able to exploit this information, as they were optimised for 
    parameters (like MFCC) obtained through systematic analysis.

%
    In pattern recognition research, it is well known that
     signal analysis and recognition are modules that are closely related.
     For example, very good matching
     between parameter vector (such as MFCC) distributions and 
     recognition models (such as HMM) yields better performance than 
     systems using auditory cues but with mismatched pattern 
     recognisers.
    Further discussion is given by M. Hunt
     in~\cite{hunt1989,hunt1999}. 
     
     Since then, research in neuroscience and auditory perception has
	advanced, yielding greater understanding  of the auditory system
	along with more sophisticated tools for the recognition of time-organised 
	features. See for example the work by Zotkin \textit{et al.}~\cite{zotkin2003}.
	
%

 \subsection{Speech Recognition with Ensemble Interval Histograms}
 Oded Ghitza  proposed in 1994 and 1995 the use of an auditory 
peripheral model for speech recognition~\cite{ghitza1994,sandhu1995} 
that simulates a great number of neurones with different internal 
threshold values. O.~Ghitza introduced the notion of the \emph{Ensemble Interval 
Histograms} representation
(EIH). That representation carries information about
the spiking time interval 
distributions
 from a population of primary auditory fibres. 
Experiments were made on the TIMIT database by using a mixture of 
Gaussian Hidden Markov Models. He observed that the EIH representation is 
more robust on distorted speech when compared to MFCC. On clean 
speech there was no gain in using that model.

It is important to note that EIH  carries information on 
averaged spiking intervals, thus specific sequences of spikes can not be 
identified inside a population of neurones. Furthermore, the 
representation has to be smoothed to be compatible with the use of a
conventional fixed frame pattern recogniser (HMM with multi-Gaussian). 
Therefore, fine grained information is lost.  

We suggest to use a similar front-end as proposed by Ghitza,
but 
to
 preserve the time structure organisation of 
spiking sequences across neurones, without computing the histograms. 
As it prevents the conventional use of HMM, we examine 
potential techniques to recognise specific spiking sequences. 
Different  
coding schemes can be used to perform the recognition.
In a collaborative work~\footnote{S.~Loiselle has been a visiting 
student in 
CERCO, Toulouse, France (Simon Thorpe and 
 Daniel Pressnitzer) during his 
 2003 summer session.}, the Rank Order Coding scheme is explored. The 
 ROC scheme has been proposed for 
visual categorisation by Thorpe \textit{et al.}~\cite{thorpe1996,thorpe2001}.

The peripheral auditory
system is crudely modelised and it is assumed
that the auditory image representation can be
processed as images. From preliminary experiments~\cite{loiselle2004}, 
it is  observed that 
 bio-inspired approaches 
     could be a good complement to statistical speech recognisers as they 
     might reach very 
     quickly acceptable results on very limited training sets. 
 \section{Conclusion}
 Conventional speech analysis and recognition techniques can yield 
 good performance levels when correctly trained and when the 
 test 
 conditions match those of the training set. But for real-life 
 situations, the designer has to train the system on huge databases that are 
 very costly to implement. On the other hand, bio-inspired processing 
 schemes can be 
 unsupervised and generalise relatively well from limited data. They 
 could efficiently complement conventional speech processing and 
 recognition techniques. Due to the intrinsic spiking nature of 
 neurones, suitable signal representations have to be found to 
 adequately adapt the signal information to the neural networks.



 \end{document}